\input{aipcheck}

\documentclass[
    ,final            
  ]
  {aipproc}

\layoutstyle{8x11single}

\begin{document}

\title{Exploring the role of model parameters and regularization procedures in the thermodynamics of
  the PNJL model}

\classification{11.30.Rd, 11.55.Fv, 14.40.Aq} \keywords {Polyakov--Nambu--Jona-Lasinio
model, equation of state, isentropic trajectories}

\author{M. C. Ruivo}{
  address={Centro de F\'{\i}sica Computacional, Departamento de F\'{\i}sica, Universidade
de Coimbra, P-3004-516 Coimbra, Portugal} }

\author{ P. Costa}{
  address={Centro de F\'{\i}sica Computacional, Departamento de F\'{\i}sica, Universidade
de Coimbra, P-3004-516 Coimbra, Portugal} }

\author{H. Hansen}{
  address={IPNL, Universit\'e de Lyon/Universit\'e Lyon 1, CNRS/IN2P3, 4 rue E.Fermi,
F-69622 Villeurbanne Cedex, France} }

\author{C. A. de Sousa}{
  address={Centro de F\'{\i}sica Computacional, Departamento de F\'{\i}sica, Universidade
de Coimbra, P-3004-516 Coimbra, Portugal} }

\begin{abstract}
 The equation of
state and the critical behavior around the critical end point are studied in the context
of the Polyakov--Nambu--Jona--Lasinio  model. We prove that a convenient choice of
the model parameters is crucial to get the correct description of isentropic
trajectories. The physical relevance of the effects of the regularization procedure is
insured by   the agreement with general thermodynamic requirements. The results
are compared with simple thermodynamic expectations and lattice data.
\end{abstract}

\maketitle


\section{Introduction} 

Understanding the   QCD phase diagram in the  $(T,\mu$)--plane is of
central importance in theoretical and experimental heavy-ion physics. Various results
from QCD inspired models indicate that at low temperatures the transition may be first
order for large values of the chemical potential; on the contrary a crossover is found
for small chemical potential and large temperature.
This suggests that the first order transition line may end when the temperature
increases, the phase diagram thus exhibiting a (second order) critical endpoint (CEP)
 that can be detected
 by a new generation of experiments with
relativistic nuclei, as the CBM experiment  (FAIR) at GSI.
Fodor and Katz \cite{Karsch} claim the values $T^{CEP}=162$ MeV and $\mu^{CEP}=360$  MeV
for such a critical point, although its precise location is still a matter of debate.

As a reliable model that can treat both the chiral and the deconfinement phase
transitions, we can consider the Polyakov loop extended Nambu--Jona--Lasinio (PNJL) model
\cite{Megias:2006PRD,Ratti:2005PRD}. In the PNJL model the deconfinement phase transition
is described by the Polyakov loop. This improved field theoretical model is fundamental
for interpreting the lattice QCD results and extrapolating into regions not yet
accessible to lattice simulations. A non trivial question in NJL type models is the
choice of the parameter set and of the regularization procedure. In fact, one should keep
in mind that this type of models  are used not only to describe physical observables in
the vacuum but also to explore the physics at finite temperature and chemical potential.
As it is well known, the order of the phase transition on the axis of the
$(T,\mu)$--plane is sensitive to the values of the parameters, most notably to the value
of the ultraviolet cutoff needed to regularize the integrals. In the pure NJL model a
large cutoff leads to a second order transition, a small cutoff to a first order one
\cite{Klevansky}. An interesting feature to be noticed is that the requirement of global
accordance with physical spectrum is obtained with values of the cutoff for which the
transition is first order on the $T=0$ axis and a smooth crossover on the $\mu=0$ axis of
the phase diagram. However, it has also been shown  that different parameter sets,
although providing  a reasonable fit of hadronic vacuum observables and predicting a
first order phase transition, will lead to different physical scenarios at finite $T$ and
$\mu$ \cite{Buballa:2004PR,Costa:2003PRC}. For instance, the absolute stability of the
vacuum state at $T=0$ is not always insured.

The consequences of the choice of the parameter set for the scenarios in the
$(T,\mu)$-plane have not been discussed in the framework of the PNJL model, where the
most popular parameter set does not allow for the absolute stability of the vacuum at
$T=0$.
In the present work, our main goal is to  analyze this problem and we will assume that
the most reliable parametrization of both NJL and PNJL models positively predicts the
existence of the CEP in the phase diagram, together with the formation of stable quark
droplets in the vacuum state at $T=0$.
Concerning the  regularization of some integrals, since, as it has been noticed by
several authors, the three dimensional cutoff is only necessary at zero temperature, the
dropping of this cutoff at finite $T$ is also carefully analyzed in this work.


\section{Model Lagrangian}

The generalized Lagrangian of the  quark model for $N_f=2$ light quarks and
$N_c=3$ color degrees of freedom,   where the quarks are coupled to a (spatially
constant) temporal background gauge field (represented in term of Polyakov loops), is
given by \cite{Ratti:2005PRD}:
\begin{eqnarray}
{\mathcal L_{PNJL}\,}\,=\,\bar q\,(\,i\, {\gamma}^{\mu}\,D_\mu\,-\,\hat m)\,q +
\frac{1}{2}\,g\,[\,{(\,\bar q\,q\,)} ^2\,\,+\,\,{(\,\bar q
\,i\,\gamma_5\,\vec{\tau}q\,)}^2\,]\,-\, \mathcal{U}\left(\Phi[A],\bar\Phi[A];T\right),
\label{eq:lag}
\end{eqnarray}
where the quark fields $q\,=\,(u,d)$ are defined in Dirac and color spaces, and
$\hat{m}=\mbox{diag}(m_u,m_d)$ is the current quark mass matrix. The pure NJL sector
contains three parameters: the coupling constant $g$, the cutoff $\Lambda$ and the
current quark mass $m=m_u=m_d$, to be determined  by fitting the experimental
values of several physical quantities (see Table~1).

The quarks are coupled to the gauge sector {\it via} the covariant derivative
$D^{\mu}=\partial^\mu-i A^\mu$. The strong coupling constant $g_{Strong}$ has been
absorbed in the definition of $A^\mu$.

The Polyakov loop $\Phi$ (the order parameter of $Z_3$ symmetric/broken phase transition
in pure gauge) is the trace of the Polyakov line defined by:
$ \Phi = \frac 1 {N_c} {\langle\langle \mathcal{P}\exp i\int_{0}^{\beta}d\tau\,
    A_4\left(\vec{x},\tau\right)\ \rangle\rangle}_\beta$,
 where ${\langle\langle  \ldots \rangle\rangle}_\beta$ with $\beta = 1/T$ is the thermal
expectation value in the grand canonical ensemble.

\newcommand{\quarkdensity}{$|\langle{\bar \psi}_u\psi_u\rangle|^{1/3}$}
\vskip0.3cm
\begin{table}[h]
        \begin{tabular}{cccccccc}
          \hline
            $ $ & $\Lambda$ &  $g$   &  $m$ &\phantom{\bigg(}\quarkdensity\ & $f_{\pi}$ & $m_{\pi}$ & $M$ \\
            $ $ & [GeV] & [GeV$^{-2}$] & [MeV] & [MeV] & [MeV] & [MeV] & [MeV]\\
            \hline
            Set A$  $ &\phantom{\bigg(} 0.590 &               7.0 &           6.0 &
                   241.5 &               92.6 &         140.2 & 400\\

            Set B$  $ &\phantom{\bigg(} 0.651 &               5.04 &           5.5 &
                   251 &               92.3 &         139.3 & 335\\
   \hline
        \end{tabular}
    \caption{Set of parameters ($\Lambda,\,g,\,m$) used in the  NJL sector of the PNJL model and the
     physical quantities chosen to fix the parameters.}
    \label{table:paramNJL}
\end{table}
\vskip0.3cm

The pure gauge sector is described by an effective potential
$\mathcal{U}\left(\Phi[A],\bar\Phi[A];T\right)$ that takes the form

\begin{eqnarray}
    \frac{\mathcal{U}\left(\Phi,\bar\Phi;T\right)}{T^4}
    &=&-\frac{b_2\left(T\right)}{2}\bar\Phi \Phi-
    \frac{b_3}{6}\left(\Phi^3+
    {\bar\Phi}^3\right)+ \frac{b_4}{4}\left(\bar\Phi \Phi\right)^2\;,
    \label{Ueff}
\end{eqnarray}%
where
\begin{eqnarray}
    b_2\left(T\right)&=&a_0+a_1\left(\frac{T_0}{T}\right)+a_2\left(\frac{T_0}{T}
    \right)^2+a_3\left(\frac{T_0}{T}\right)^3.
\end{eqnarray}


The coefficients   $a_i$ and $b_i$ of the Polyakov loop effective potential are
chosen (see \cite{Ratti:2005PRD}) to reproduce, at the mean-field level, the results
obtained in lattice calculations. The numerical values are: $a_0=6.75$, $a_1=-1.95$, 
$a_2=2.625$, $a_3=-7.44$, $b_3=0.75$ and $b_4=7.5$. In addition, we choose  $T_0=190$ MeV  
at finite temperature and $T_0=270$ MeV at finite temperature and chemical potential.

The important point of our argumentation about the choice of the model parameters comes
from the comparison between  the point $(0,\mu_{c})$ of the phase diagram, where
$\mu_{c}$ is the position of the first order line at zero temperature, and the point
$(0,M_{vac})$, where $M_{vac} = M$ is the mass of the $u,d$--quark in the vacuum.
Two special cases are observed \cite{Buballa:2004PR}:
\begin{itemize}
  \item [(i)] For set A, the first order phase transition occurs at $\mu_{c}$ such that 
   $\mu_{c}<M_{vac}$, and  consequently  the phase transition connects the vacuum state 
  ($\rho_{q}=0$) directly with the phase of partially restored chiral symmetry 
  ($\rho_{q}=\rho_{c}$). This is compatible with the existence of stable quark matter, 
  indicating the possibility for finite droplets to be in mechanical equilibrium with the 
  vacuum at zero pressure \cite{Buballa:2004PR,Costa:2003PRC}.
  \item [(ii)]  For set B,  $\mu_{c}> M_{vac}$, so  the phase transition connects a 
  $\rho_{q}\neq 0$  phase of massive quarks with  the phase of partially restored chiral 
  symmetry ($\rho_{q}=\rho_{c}$). From the physical point of view, this scenario is 
  unrealistic because it predicts the existence of a low-density phase of homogeneously 
  distributed constituent quarks \cite{Buballa:2004PR}.
\end{itemize}

So, although we can choose several sets of parameters which fit physical observables in
the vacuum, we notice, however, that the value of the cutoff itself does have some impact
on the characteristic of the first order phase transition. Comparing  the two sets of
parameters A and B (see Table 1) we verify that for larger values of the cutoff, as in set B, 
a more strong attraction is necessary both to reproduce the physical values in the vacuum 
and to insure a  first order phase transition. The more realistic choice
is provided by set A with important  implications  on the reliability of 
isentropic trajectories as we will discuss in the sequel.

Concerning the  regularization of some integrals, since the three dimensional cutoff is
only necessary at zero temperature, the dropping of this cutoff at finite $T$ is also
considered. This procedure allows for the presence of high momentum quark states, leading
to interesting physical consequences, as it has been shown in \cite{Costa:2008PRD2},
where the advantages and drawbacks of this regularization have been discussed in the
context of the NJL model. We will  enlarge the use  this procedure to the PNJL model and
discuss its influence on the behavior of several relevant observables.
So, we will consider  two different regularization procedures at $T\neq 0$:
\begin{itemize}
  \item [(i)]{\em Case I}, where the cutoff is used  only in the integrals
that are divergent ($\Lambda \rightarrow \infty$ in the convergent ones).
\item [(ii)]{\em Case II},
where the  regularization consists in the use of the cutoff $\Lambda$ in all integrals.
\end{itemize}

Let us notice that the choice of a regularization procedure is a part of the effective
modeling of QCD thermodynamic. Indeed  we found that a comprehensive study of the
differences between the two regularization procedures (with and without cutoff on the
quark  momentum states at finite temperature) is necessary to have a better understanding
of the PNJL model and the role of high momentum quarks around the phase transition.
The physical relevance of our numerical solutions is insured by demanding the agreement
with general thermodynamic requirements.  In particular, we   will verify that the
correct description of isentropic lines is closely related with the parameter choice in
the pure NJL sector.


\section{Numerical results and conclusions}
\begin{figure}[h]
\vspace{-2cm}
\hspace{0.35cm}\includegraphics[width=0.55\textwidth]{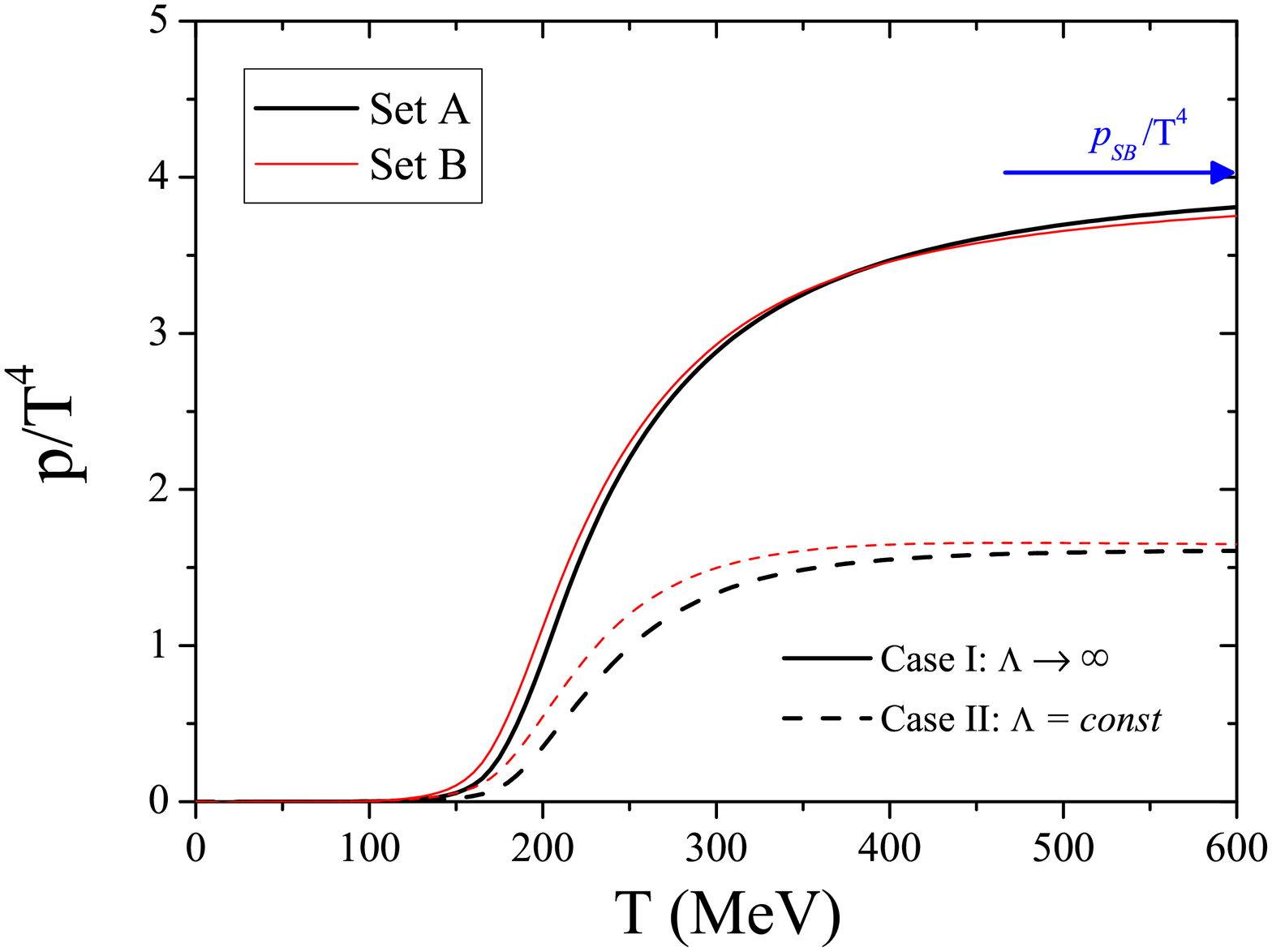}
\hspace{-0.85cm}\includegraphics[width=0.55\textwidth]{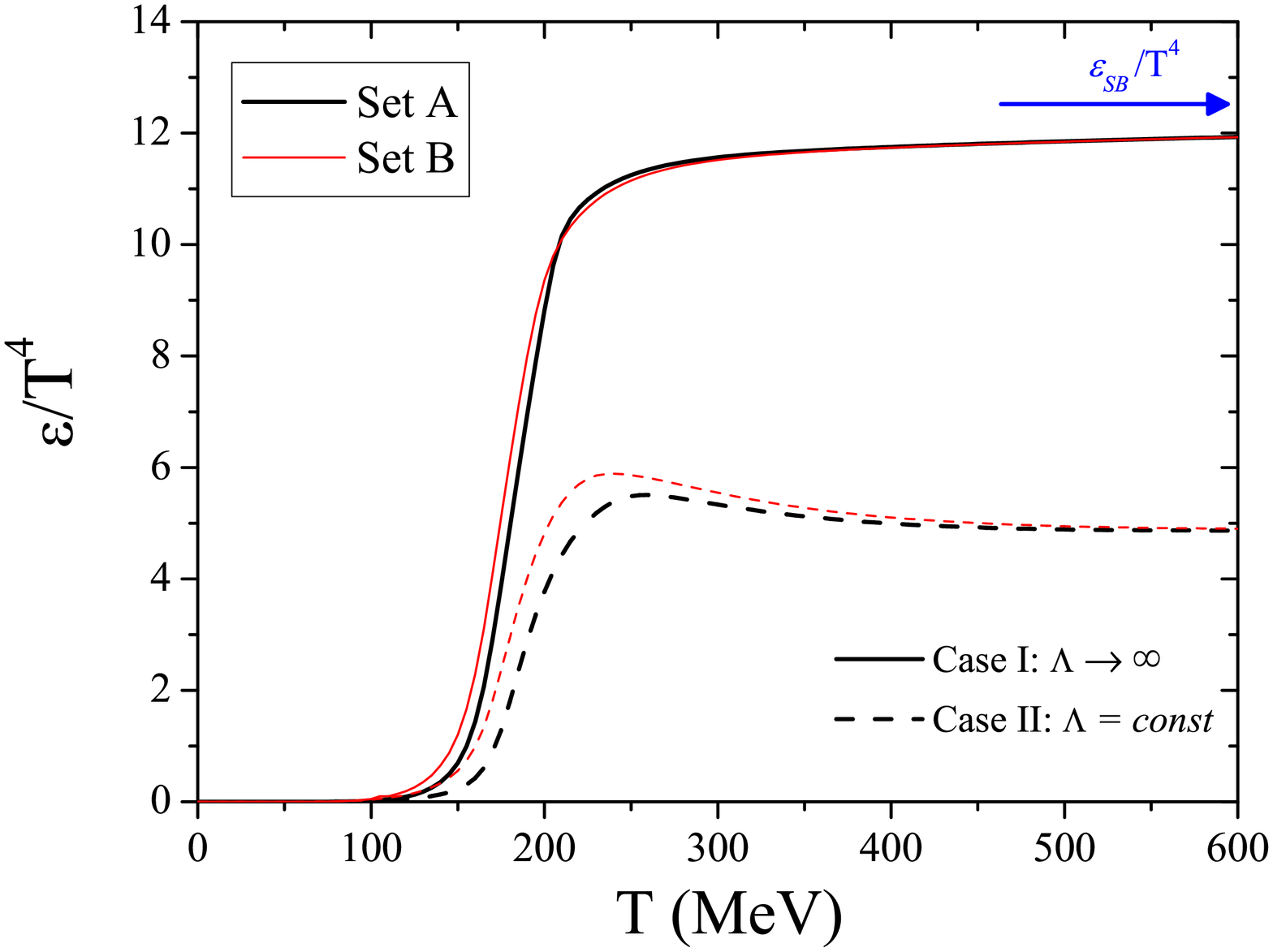}
\caption{ Scaled pressure $(p)$ and   energy  per particle $(\epsilon)$ as a function 
of the temperature at zero chemical potential for both sets
of parameters A and B, and both regularization procedures.} \label{Fig:1}
\end{figure}

To  check the usefulness of the regularization procedure  we start by considering our
numerical results at vanishing quark chemical potential. To this purpose, we plot   the
scaled pressure and the energy  as functions of the temperature in Fig. \ref{Fig:1}.
%
The transition to the high temperature phase is a rapid crossover rather than a phase
transition and, consequently, the pressure and the energy densities, which have a 
similar behavior,  are continuous functions of the temperature.

We verify that the  inclusion of the Polyakov loop together with the regularization procedure
implemented in case I, is essential to obtain the required increase of extensive
thermodynamic quantities, insuring the convergence to the Stefan--Boltzmann (SB) limit of
QCD. Some comments are in order concerning the role of the regularization
procedure for $T>T_c$, where $T_c$ is the reduced temperature. In this temperature range, 
due to the presence of high momentum
quarks, the physical situation is dominated by the significative decrease of the
constituent quark masses by the $q \bar q$ interactions.  This allows for an ideal gas
behavior of almost massless quarks with the correct number of degrees of freedom. 
In this context both sets of parameters provide similar conclusions as can be seen in Figure 1.

The advantage of our phenomenological model is the possibility to provide equations of
state at nonzero chemical potential too \cite{Costa:2009new}. So, we can also test its 
ability to reproduce~recent progress in lattice QCD with small non vanishing chemical potential.

\vskip0.3cm

The isentropic lines play a very important role in the understanding of thermodynamic
properties of matter created in relativistic heavy ion collisions.  Some model calculations predict that  in a
region around the CEP the properties of matter are only slowly modified as the collision
energy is changed, as a consequence of the attractor character of the CEP
\cite{Stephanov:1998PRL}.

\begin{figure}[t]
\vspace{-2cm}
\hspace{0.35cm}\includegraphics[width=0.55\textwidth]{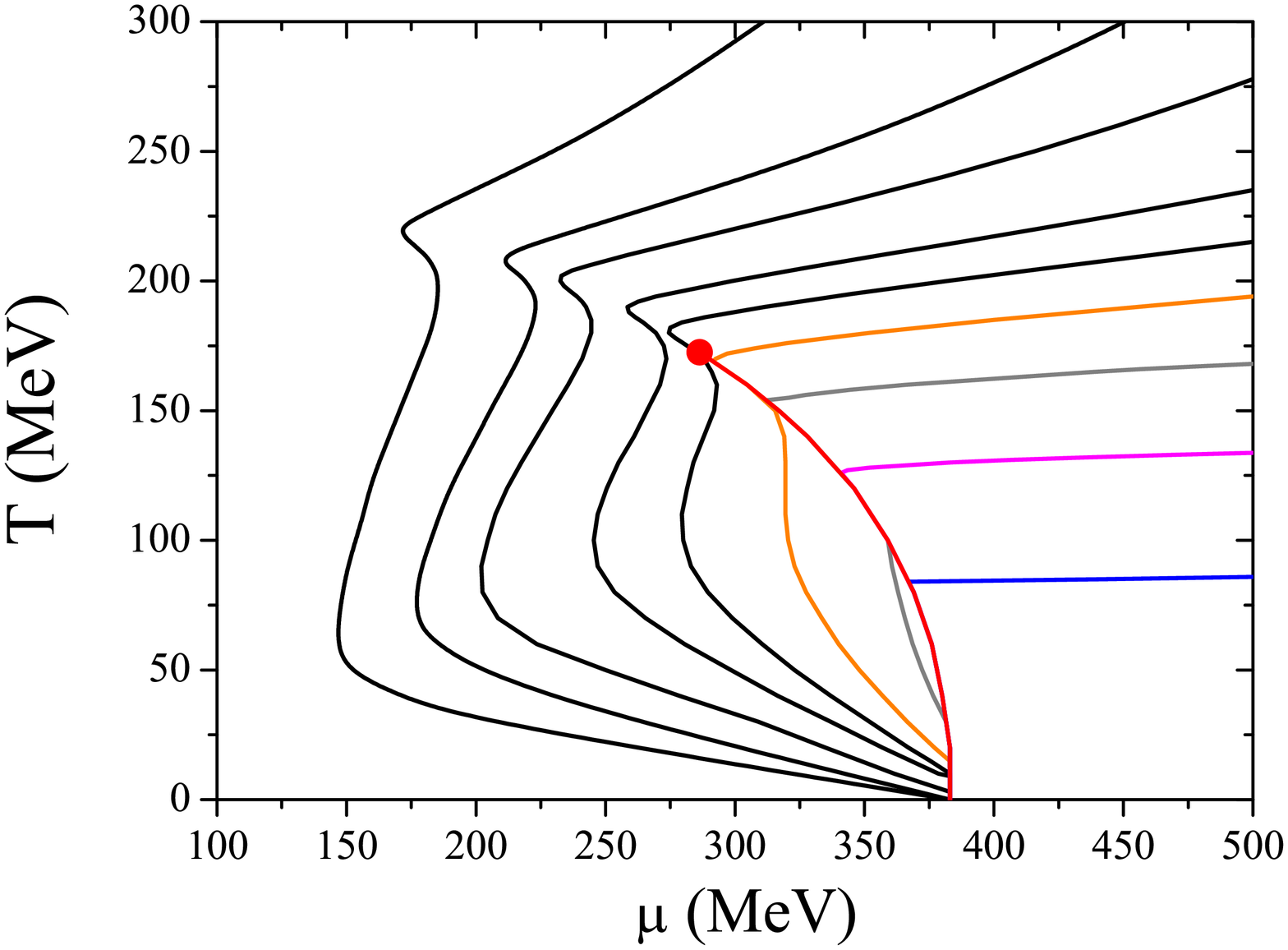}
\hspace{-0.85cm}\includegraphics[width=0.55\textwidth]{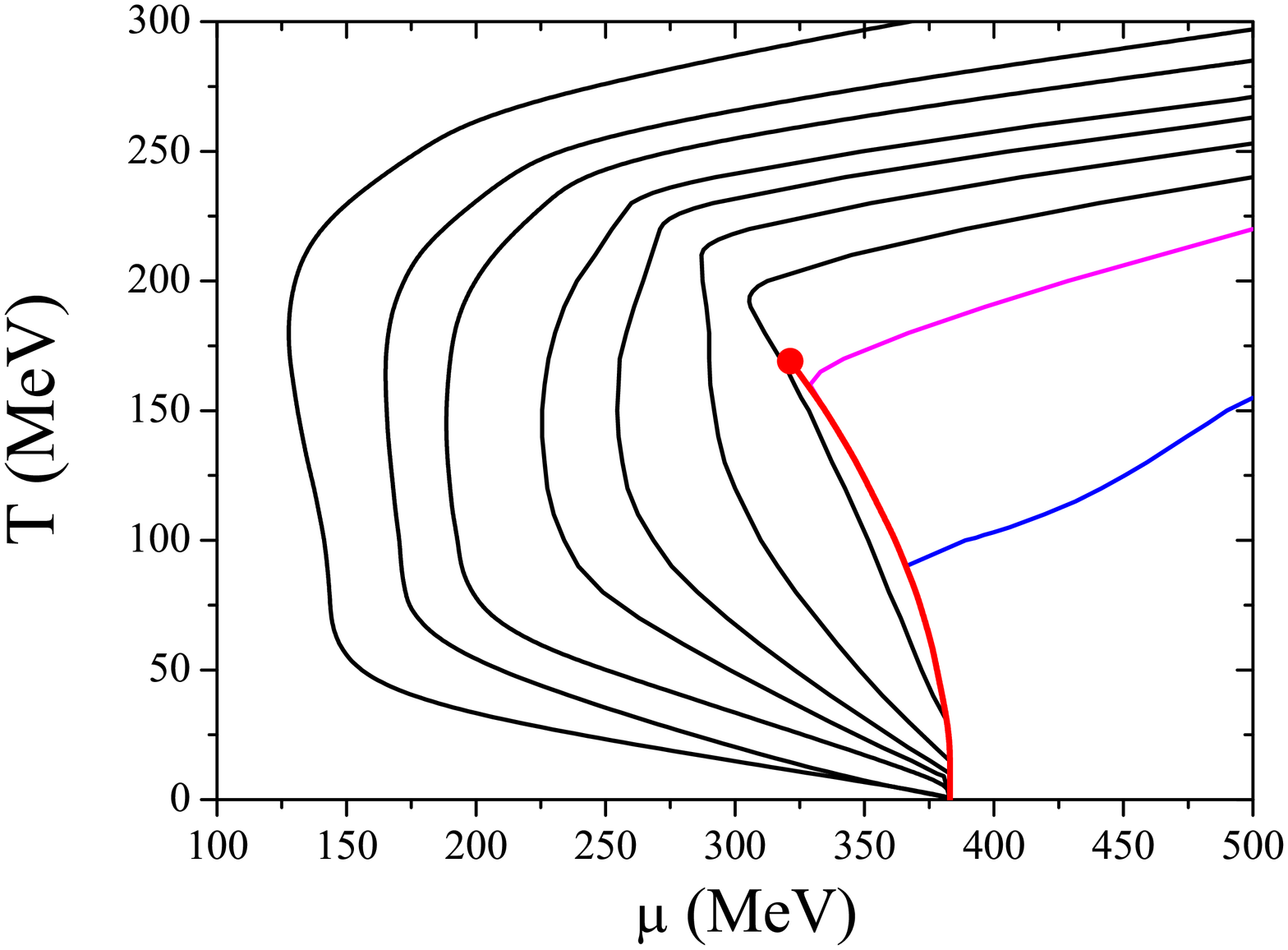}
\caption{ Isentropic trajectories  in the $(T,\mu)$-plane for case I (left
panel) and case II (right panel) using the parameter set A. The following values of the
entropy per baryon number have been  considered:
$s/\rho_q=1,\,2,\,3,\,4,\,5,\,6,\,8,\,10,\,15$ (anticlockwise direction). }
\label{Fig:4}
\end{figure}

Our numerical results  for the isentropic lines in the $(T,\mu)$-plane are shown in Fig.
\ref{Fig:4}, where we have used  set A of parameters and both regularization procedures.

We start the discussion  by analyzing the behavior of the isentropic lines in the limit
$T\rightarrow 0$.  We point out that, as already referred by other authors
\cite{Scavenius}, in this limit: (i) $s \rightarrow 0$, according to the third law of thermodynamics; and (ii) for $s/\rho_q\,=\,const$, we have to insure that also $\rho_q \rightarrow 0$.
However, the satisfaction of the condition (ii) is only provided when $\mu\leq M_{vac}$ as discussed above.
In spite of the general use of set B in the literature of the PNJL model, only set A
satisfies  this ansatz.
We remember  that with  set A we are, at $T=0$, in the presence of droplets
(states in mechanical equilibrium with the vacuum state at $P=0$).
Consequently,  all isentropic trajectories
directly terminate at the first order transition line at $T=0$. So, for set A it is
verified that $s \rightarrow 0$ and $\rho_q \rightarrow 0$  in the limit $T\rightarrow
0$,  as it should be.

In conclusion, our convenient choice of the model parameters allows a first order phase
transition that is stronger than in other treatments of the NJL (PNJL) model. This choice
is crucial to obtain important results:  the criterium of stability of the quark droplets
\cite{Buballa:2004PR,Costa:2003PRC} is fulfilled, and, in addition,  simple thermodynamic
expectations in the limit $T\rightarrow 0$ are verified.

At $T\neq0$, in the first order line, the behavior we find is somewhat different from
those claimed by other authors \cite{Nonaka:2005PRC} where a phenomena
of focusing of trajectories towards the CEP is observed. For case I (see Figure
\ref{Fig:4}, left panel) we see that the isentropic lines with $s/\rho_q=1,...,4$ come
from the region of symmetry partially restored and attain directly the phase transition,
going along with the phase transition as $T$ decreases until it reaches $T=0$. The same
behavior is found for case II when $s/\rho_q=1,2$ (see Figure \ref{Fig:4}, right panel).
For case II, we also observe, in a small range of $s/\rho_q$ around~$3$, a tendency to
convergence of these isentropic lines towards the CEP. These lines come from the region
of symmetry partially restored in the direction of the crossover line. For smaller values
of $s/\rho_q$, the isentropic lines turn about the CEP and then attain the first order
transition line. For larger values of $s/\rho_q$ the isentropic trajectories approach the
CEP by the region where the chiral symmetry is still broken, and also attain the first
order transition line after bending toward the critical point.
As already pointed out in \cite{Scavenius}, this is a natural result in these type of
quark models with no change in the number of degrees of freedom of the system in the two
phases. As the temperature decreases a first order phase transition occurs, the latent
heat increases and  the formation of the mixed phase is thermodynamically favored.

In the crossover region, for both cases, the behavior of the isentropic lines is
qualitatively similar to the one obtained in lattice calculations \cite{Ejiri:2006PRD} or
in some models \cite{Nonaka:2005PRC}.
On the other hand, the isentropic trajectories in the phase diagram indicate that the
slope of the trajectories goes to large values for large $T$.

Summarizing, we have considered the  PNJL model as one of the prototype models of dynamical symmetry
breaking  of QCD  and investigated the phase
structure at finite $T$ and $\mu$.  Working out of the chiral limit, a CEP which
separates the first and the crossover line is found.
Two important points of our model calculation concern the  choice of the model parameters with the emphasis on the parameter choice  which is compatible with the formation of
stable droplets at zero temperature.
The effects of two regularization procedures at finite temperature, one that allows high
momentum quark states to be present (case I) and the other not (case II), have also been discussed.
We conclude that the choice of the model parameters has  important consequences in order
to obtain the correct asymptotic low temperature behavior. In the zero temperature limit,
the chemical potential approaches a finite value that must satisfy to the condition
$\mu_c<M_{vac}$. Only the set of parameters  A  insures this condition that allows us to
obtain both $s=0$ and $\rho_q=0$.
In addition, the regularization procedure is important for obtaining agreement with the asymptotic
behavior above $T_c$.

\begin{theacknowledgments}
Work supported by FCT under project CERN/FP/83644/2008.
\end{theacknowledgments}

\bibliographystyle{aipproc}   


\IfFileExists{\jobname.bbl}{}
 {\typeout{}
  \typeout{******************************************}
  \typeout{** Please run "bibtex \jobname" to optain}
  \typeout{** the bibliography and then re-run LaTeX}
  \typeout{** twice to fix the references!}
  \typeout{******************************************}
  \typeout{}
 }

\end{document}